# UNIQUE FUTURES IN CHINA: STUDYS ON VOLATILITY SPILLOVER EFFECTS OF FERROUS METAL FUTURES

Tingting Cao[1,2,3], Weiqing Sun*[1,4], Cuiping Sun[1], Lin Hao[1]

[1]Department of Economics and Management, Weifang University of Science and Technology, 1299 Jinguang Street, Shouguang, Shandong, China
[2]Postdoctoral Research Station of Weihai City Commercial Bank, 11666 Jingshidong Road, Jinan, China
[3]Postdoctoral Research Station of Center for Economic Research Shandong University, 27 Shanda Nanlu, Jinan, China
[4]Department of Economics and Management, The University of Suwon, 17, Wauan-gil, Bongdam-eup, Hwaseong-si, Gyeonggi-do 18323, Republic of Korea

**Abstract:** Ferrous metal futures have become unique commodity futures with Chinese characteristics. Due to the late listing time, it has received less attention from scholars. Our research focuses on the volatility spillover effects, defined as the intensity of price volatility in financial instruments. We use DCC-GARCH, BEKK-GARCH, and DY(2012) index methods to conduct empirical tests on the volatility spillover effects of the Chinese ferrous metal futures market and other parts of the Chinese commodity futures market, as well as industries related to the steel industry chain in stock markets. It can be seen that there is a close volatility spillover relationship between ferrous metal futures and nonferrous metal futures. Energy futures and chemical futures have a significant transmission effect on the fluctuations of ferrous metals. In addition, ferrous metal futures have a significant spillover effect on the stock index of the steel industry, real estate industry, building materials industry, machinery equipment industry, and household appliance industry. Studying the volatility spillover effect of the ferrous metal futures market can reveal the operating laws of this field and provide ideas and theoretical references for investors to hedge their risks. It shows that the ferrous metal futures market has an essential role as a "barometer" for the Chinese commodity futures market and the stock market.

**JEL Classification:** G13; G14; G15.

**Keywords:** volatility spillover; Spillover effect; ferrous metal; BEKK-GARCH; DCC-GARCH; DY(2012) index

## 1 Introduction

Ferrous metal is "iron, manganese and chromium, and their alloys." Steel and its processing raw materials such as iron ore, coke, and ferroalloys all belong to the category of ferrous metals. Since the rebar futures were listed on the Shanghai Futures Exchange in March 2009, China has successively listed important ferrous metal commodity futures, such as wire, hot-rolled coils, iron ore, coke, ferrosilicon and silicon-manganese. The trading volume of ferrous metals futures is huge, relying on the solid strength of China's steel manufacturing industry and the massive demand for steel. At present, it has accounted for more than a quarter of China's commodity futures trading volume and has become a unique commodity future with Chinese characteristics.

The Autoregressive Conditional Heteroscedasticity (ARCH) model was first proposed by Engle(1982). This model describes the fluctuation of financial market returns and the heteroscedasticity of time series. Based on this, Bollerslev(1986) developed the univariate generalized autoregressive conditional heteroscedasticity (GARCH) model. The univariate GARCH model captures characteristics such as fluctuation aggregation, conditional heteroscedasticity, "spikes" and "thick tails" of high-frequency financial time series. Bollerslev et al. (1988) considered using the multivariate GARCH model to study the volatility spillover relationship between multivariable. The multivariate GARCH model can accurately describe the vertical agglomeration effect of the prices of multiple financial assets along the time direction and effectively capture the horizontal risk cross-transfer effect between the prices of different financial assets.

Based on the original multivariate GARCH, scholars have continuously improved and optimized the parameter estimation of the conditional covariance matrix of random error terms and proposed VEC-GARCH, BEKK-GARCH, DCC-GARCH, and other optimized GARCH models. Lien (2009) uses the DCC-GARCH model to empirically test the short-term return and volatility spillover relationship between the futures copper prices of the three trading markets of the London Metal Exchange, the New York Mercantile Exchange, and the Shanghai Futures Exchange. The results show that there exists a mutual spillover relationship between each of the two markets. Xing et al. (2011) used a multivariate T-GARCH model to study the volatility spillover effect between China's stock index futures and spot market and find a significant mutual volatility spillover effect between the two markets. Liu et al. (2012) used the BEKK-MGARCH model to study the volatility spillover effect between the CSI 300 simulated stock index futures and the spot market. The results show a significant mutual spillover effect. Kang et al. (2013) found a strong volatility relationship between futures and spot markets in the Korean stock market using the BEKK-GARCH model. Sogiakas and Karathanassis (2015) used the Dynamic Conditional Correlation GARCH model (DCC-GARCH) to investigate the direction and causality of the spillover effects between cash and derivatives markets.

Alotaibi and Mishra (2015) examined the effects of return spillovers from Saudi Arabia and U.S. markets to GCC stock markets by the bivariate BEKK-GARCH model, and significant return spillover effects exist from Saudi Arabia and the U.S. to GCC markets. Baldi et al.(2016) found a negative volatility transmission relationship between SP500 and the commodities market through the BEKK-GARCH model. Wang and Li (2016) empirically analyzed the impact of the night trading system on the linkage relationship between China and the U.S. futures market through a multivariate GARCH model. Fu et al. (2017) used the model of VEC-BEKK-GARCH and DCC-MGARCH to study the price linkage effect of the continuous trading system on gold futures between Chinese and U.S. futures markets. Huo and Ahmed (2017) observed a leading role of the Shanghai stock market in the Hong Kong stock market in terms of both mean and volatility spillover effects after the Stock Connect.

Roy (2017) applies the DCC-MGARCH model to the daily commodity futures price index. The financial contagion and volatility spillover effects between Indian commodity derivative market bond, foreign exchange, gold, and stock markets are studied. Zheng and Ma (2018) use the DCC-GARCH model and BEKK-GARCH model to investigate the relationship and spillover effects between China's egg futures and the spot market. Dieijen et al.(2018) investigate whether volatility in user-generated content (UGC) can spill over to volatility in stock returns and vice versa by a multivariate GARCH model. Chang et al.(2019) focus on the volatility spillover effects between the agricultural and energy industries with the BEKK-GARCH model. Li (2020) examines the volatility spillovers of interrelated European stock markets under the uncertainty of Brexit with the BEKK-GARCH model and finds that the U.K.'s influence on the other European markets has decreased since the campaign for the E.U. referendum started in January 2016. Yang (2020) studies the risk spillovers of the Chinese stock market to major East Asian stock markets during turbulent and calm periods with the MSGARCH-EVT-copula approach. The spillover index approach suggested by Diebold and Yilmaz (2009), this methodology has been further improved in Diebold and Yilmaz (2012), in which the results are invariant to variable ordering. Antonakakis et al. (2016) use an introduced spillover index to examine dynamic spillovers between spot and futures markets volatility, the volume of futures trading, and open interest in the U.K. and the U.S. Lau and Sheng (2018) examine the inter-and intra-regional spillover effects across international stock markets in London, Paris, Frankfurt, Toronto, New York, Tokyo, Shanghai, Hong Kong, and Mumbai by using both symmetric and asymmetric causality tests. Su (2019), based on the spillover index, used a regression analysis applied to G7 and BRICS stock markets, from which new insights emerged as to the excessive risk spillovers that arose in G7 and BRICS stock markets and revealed how extreme risk spillovers across developed. Kang et al.(2019) investigated the dynamic spillovers between ASEAN-5 and world stock markets using a dynamic equicorrelation (DECO) model and the spillover index. Jiang et al. (2019), the DCC-GARCH model is firstly used to study the relationship between the oil market and China's commodities. Yin et al. (2020) applied a spillover index method to investigate the interindustry volatility spillovers in Shanghai Stock Exchange from 2009 to 2018. Zhang et al. (2020) employ the BEKK-GARCH model to construct the volatility network of G20 stock markets. Chen et al. (2020) use the DCC-GARCH model to depict the dynamic correlations between rare earth and new energy markets.

This paper investigates the volatility spillovers between ferrous metal futures and other parts of commodity futures, the ferrous metal futures and the stock market. In the second chapter of this article, a literature review of scholars' research on the spillover effect of financial market volatility is carried out. In the third chapter of this article, the DCC-GARCH and BEKK-GARCH methods are introduced to make an empirical test on volatility spillover effects for ferrous metal futures. The empirical tests of DCC-GARCH and BEKK-GARCH model for the spillover relationship between ferrous metal futures and other commodity futures, ferrous metal futures, and the Chinese stock market are implemented in the fourth and fifth parts. In the sixth part, we use the Diebold and Yilmaz (2012) volatility spillover index (the DY(2012) index) to measure the extent of volatility transfer among markets.

The various sectors of the Chinese financial market are not isolated. As an essential practice of financial innovation, the futures market has important functions of risk hedging and price discovery. Studying the volatility spillover effect of the ferrous metal futures market can reveal the operating laws of this field and provide ideas and theoretical references for investors to hedge their risks.

## 2 Research methods of volatility spillover effects

GARCH model (Generalized autoregressive conditional heteroscedasticity model) affects the volatility spillover in financial markets. It is a commonly used method to study financial markets. Moreover, we use the DCC-GARCH and BEKK-GARCH methods to make an empirical test on volatility spillover effects for ferrous metal futures.

Diebold and Yilmaz (2009) developed a volatility spillover index (DY index) based on forecast error variance decomposition from vector autoregressions (VARs) to measure the extent of volatility transfer among markets. This methodology has been further improved by Diebold and Yilmaz (2012), who used a generalized VAR framework in which forecast-error variance decompositions are invariant to variable ordering. The DY index is a versatile measure allowing dynamic quantification of numerous aspects of volatility spillovers. DY addresses total spillovers (from/to each market i, to/from all other markets, added across i). It can also examine directional spillovers (from/to a particular market). Here we also use the DY(2012) index to measure the directional spillovers of ferrous metal futures to others.

### 2.1 DCC-GARCH model

Firstly, the univariate GARCH (1,1) model is established for each return sequence. The specific formulas can be set as follows:

$$R_t = \phi_0 + \sum_{i=1}^{p} \phi i R_{t-i} + \varepsilon_i \qquad (1)$$

$$\varepsilon_i = \sigma_t Z_t; \varepsilon_i \mid \Omega_{t-1} \sim N(0, \sigma_t^2) \qquad (2)$$

$$\sigma_t^2 = \omega + \alpha \varepsilon_{t-1}^2 + \beta \sigma_{t-1}^2 \qquad (3)$$

$R_t$ is the logarithmic return sequence. $\phi_0$ is the coefficient of the autoregressive term for the logarithmic return series. This part represents the part of the yield that can be predicted by the mean equation. $\varepsilon_i$ is the residual term without autocorrelation, representing the unexpected return caused by innovation. $\sigma_t^2$ is the conditional variance. $Z_t$ is the standard residual of $\varepsilon_i$. $\Omega_{t-1}$ is a collection of all historical information from the period of $t-1$. α is the coefficient of the residual. Its lag is the ARCH effect's coefficient and represents the volatility's agglomeration. β is the coefficient autoregressive for conditional variance, which is the coefficient for the GARCH effect and represents the persistence of volatility. The closer the sum of α and β is to 1, it indicates that the fluctuation of the current period has a more significant influence on the fluctuation of the next period. The DCC-MGARCH (1,1) model should be constructed to study the volatility spillover effect between ferrous metal futures and other commodity futures in the Chinese futures market and between ferrous metal futures and the relevant sections of the Chinese stock market. The specific settings of the model are as follows:

According to the above definition, the correlation coefficient of the residuals for the mean equations between every two markets can be expressed as:

$$\rho_{ij,t} = \frac{E(\varepsilon_{i,t}\varepsilon_{j,t})}{\sqrt{E(\varepsilon_{i,t}^2)E(\varepsilon_{j,t}^2)}} = \frac{q_{ij,t}}{\sqrt{q_{ii,t}q_{jj,t}}} \tag{4}$$

In formula (4), $\rho_{ij,t}$ is the correlation coefficient between $\varepsilon_{i,t}$ and $\varepsilon_{j,t}$ in the t-time period, and the mean value of the variable is zero; which can be expressed as:

$$q_{ij,t} = \bar{q}_{ij} + \theta(z_{i,t-1}Z_{j,t-1} - \bar{q}_{ij}) + \eta(q_{ij,t-1} - \bar{q}_{ij}) \tag{5}$$

In formula (5), $\bar{q}_{ij}$ is the unconditional variance. We define that:

$$Q_t = (q_{ij,t})_{2*2}; Q_t^* = \text{diag}(\sqrt{q_{11,t}}\sqrt{q_{22,t}}) \tag{6}$$

The dynamic correlation matrix is:

$$R_t = (\rho_{ij,t})_{2*2} = Q_t^{*-1}Q_tQ_t^{*-1} \tag{7}$$

The conditional variance covariance matrix between every two markets can be expressed as:

$$H_t = (\rho_{ij,t}, \sigma_{ii,t} * \sigma_{jj,t}) = D_tR_tD_t \tag{8}$$

In formula (8), by observing the changes in the elements of the dynamic correlation matrix, we can understand the dynamic changes in the correlation between the futures and spot markets.

**2.2 BEKK-GARCH model**

$$H_t = C_0C_0^T + \sum_{i=1}^{p} A_i\varepsilon_{t-i}\varepsilon_{t-i}^T A_i^T + \sum_{j=1}^{q} B_jH_{t-j}B_j^T \tag{9}$$

The DCC-GARCH model can reflect the linkage effect between markets through the correlation coefficient. However, it cannot reflect the direction of the volatility spillover effect between the two markets. Therefore, we take the BEKK-GARCH model to research the direction of the volatility spillover effect between the two markets.

The advantages of the BECK-GARCH model mainly come from two aspects. One is that it can ensure that the covariance matrix is positive and definite under less restrictive conditions. Moreover, fewer parameters need to be estimated in the BEKK-GARCH model. The model can better retain the information in the variance-covariance matrix between the residual sequences, effectively reflecting the direction of volatility spillover effects between markets. Compared with the model of DCC-GARCH, the BEKK-GARCH model obeys the following transformation for the conditional variance-covariance matrix of the residual vector.

In formula (9), $C_0$ represents the lower triangular matrix. $A_i$ and $B_j$ are square matrices. Now we, respectively, define i=1, 2; j=1, 2. And they represent two different markets. The matrix expressions for the BEKK-GARCH (1,1) model of the two variables are:

$$H_t = \begin{pmatrix} h_{11,t} & h_{12,t} \\ h_{21,t} & h_{22,t} \end{pmatrix}, C_0 = \begin{pmatrix} c_{11} & 0 \\ c_{21} & c_{22} \end{pmatrix}, A = \begin{pmatrix} a_{11} & a_{12} \\ a_{21} & a_{22} \end{pmatrix}, B = \begin{pmatrix} b_{11} & b_{12} \\ b_{21} & b_{22} \end{pmatrix}$$

$$(\varepsilon_{t-1}\varepsilon_{t-1}^T) = \begin{pmatrix} \varepsilon_{1,t-1}^2 & \varepsilon_{1,t-1}\varepsilon_{2,t-1} \\ \varepsilon_{2,t-1}\varepsilon_{1,t-1} & \varepsilon_{2,t-1}^2 \end{pmatrix}, H_{t-1} = \begin{pmatrix} h_{11,t-1} & h_{12,t-1} \\ h_{21,t-1} & h_{22,t-1} \end{pmatrix}$$

The volatility spillover effect of sector 2 on sector 1 is mainly reflected by the coefficients $a_{21}$ and $b_{21}$. The coefficients $a_{12}$ and $b_{12}$ mainly reflect the volatility spillover effect of sector 1 on sector 2. The null hypothesis is that $a_{21}$, $b_{21}$, $a_{12}$ and $b_{12}$ are all equal to zero, indicating no mutual fluctuation overflow relationship between sector 1 and sector 2.

**2.3 DY(2012) index**

Diebold and Yilmaz (2009) developed a volatility spillover index (DY index) based on forecast error variance decomposition from vector autoregressions (VARs) to measure the extent of volatility transfer among markets. This methodology has been further improved by Diebold and Yilmaz (2012), who used a generalized VAR framework in which forecast-error variance decompositions are invariant to variable ordering. The DY index is a versatile measure allowing dynamic quantification of numerous aspects of volatility spillovers. DY addresses total spillovers (from/to each market $i$, to/from all other markets, added across $i$). It can also examine directional spillovers (from/to a particular market).

An estimator of the daily variance using daily high and low prices. For market $i$ on day t we have:

$$\tilde{\sigma}_{it}^2 = 0.361\left[\ln(P_{it}^{max}) - \ln(P_{it}^{min})\right]^2 \tag{10}$$

The annualized daily percent standard deviation (volatility) is $\hat{\sigma}_{it} = 100\sqrt{365 \cdot \tilde{\sigma}_{it}^2}$,

Denoting the KPPS H-step-ahead forecast error variance decomposition by

$\theta_{ij}^g(H)$, for $H = 1, 2, \ldots$, we have

$$\theta_{ij}^g(H) = \frac{\sigma_{ii}^{-1} \sum_{h=0}^{H-1} \left(e_i' A_h \Sigma e_j\right)^2}{\sum_{h=0}^{H-1} \left(e_i' A_h \Sigma A_h' e_i\right)} \qquad (11)$$

We normalize each entry of the variance decomposition matrix by the row sum as

$$\tilde{\theta}_{ij}^g(H) = \frac{\theta_{ij}^g(H)}{\sum_{j=1}^N \theta_{ij}^g(H)} \qquad (12)$$

We can construct a total volatility spillover index:

$$S^g(H) = \frac{\sum_{\substack{i,j=1 \\ i \neq j}}^N \tilde{\theta}_{ij}^g(H)}{N} * 100 \qquad (13)$$

We measure the directional volatility spillovers received by market i from all other markets j as:

$$S_{i\cdot}^g(H) = \frac{\sum_{\substack{j=1 \\ j \neq i}}^N \tilde{\theta}_{ij}^g(H)}{\sum_{j=1}^N \tilde{\theta}_{ij}^g(H)} \cdot 100 \qquad (14)$$

Similarly, in this fashion, we measure the directional volatility spillovers transmitted from market $i$ to all other markets $j$ as:

$$S_{\cdot i}^g(H) = \frac{\sum_{j=1}^N \tilde{\theta}_{ji}^g(H)}{\sum_{j=1}^N \tilde{\theta}_{ji}^g(H)} * 100 \qquad (15)$$

We obtain the net volatility spillover effects: from market i to all other markets j as:

$$S_i^g(H) = S_{\cdot i}^g(H) - S_{i\cdot}^g(H) \qquad (16)$$

It is also of interest to examine net pairwise volatility spillovers, which we define as:

$$S_{ij}^g(H) = \left(\frac{\tilde{\theta}_{ij}^g(H)}{\sum_{k=1}^N \tilde{\theta}_{ik}^g(H)} - \frac{\tilde{\theta}_{ji}^g(H)}{\sum_{k=1}^N \tilde{\theta}_{jk}^g(H)}\right) \cdot 100 \qquad (17)$$

## 3 Volatility spillover effects between ferrous metal futures and other sectors in Chinese commodity futures markets

### 3.1 China's steel industry chain and ferrous metal futures

The ferrous metal industry chain is the abbreviation of the entire iron and steel industry chain from raw materials to the steel smelting industry. Upstream, midstream, and downstream are three parts of the chain. Upstream uses iron ore, coking coal, coke and other raw materials in the blast furnace to cast pig iron. The midstream is based on pig iron, according to the different requirements for the quality of steel in different fields, adding ferrosilicon, manganese silicon and other iron alloy elements, forging into different raw steel products. Downstream make raw steel products of different shapes, such as rebar, hot-rolled coils, and wire. These finished steel products enter different fields, such as real estate, infrastructure, machinery and automobile manufacturing. The basic structure of the ferrous metal industry chain is shown in Figure 1.

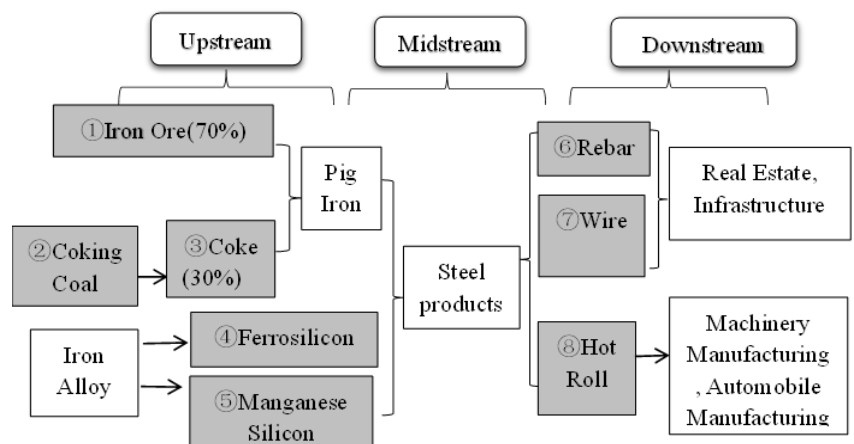

**Figure 1: Basic structure of the ferrous metal industry chain**

As can be seen from Figure 1, eight commodity futures products in the range of ferrous metals have been listed in the Chinese futures market. According to data released by the World Iron and Steel Association, the global crude steel output in 2017 reached 1.691 billion tons, an increase of 5.3% over 2016. China's crude steel output reached 832 million tons, accounting for 49.20% of the world's crude steel output, the world's largest iron and steel producing country.

The rebar futures were listed on the Shanghai Futures Exchange in March 2009. China has now listed iron ore futures, coking coal futures, coke futures, ferrosilicon futures, manganese silicon futures, rebar futures, wire futures, and hot-rolled coils. The specific time contract details of each future product listed are shown in Table 1.

**Table 1: Listing details of China's ferrous metal futures**

| Time to list on the market | Exchange | Listed varieties |
|---|---|---|
| March 27, 2009 | Shanghai Futures Exchange | Rebar futures, wire futures |
| April 15, 2011 | Dalian Commodity Exchange | Coke futures |
| March 22, 2013 | Dalian Commodity Exchange | Coking coal futures |
| October 18, 2013 | Dalian Commodity Exchange | Iron ore futures |
| March 21, 2014 | Shanghai Futures Exchange | Hot-rolled coil futures |
| August 8, 2014 | Zhengzhou Commodity Exchange | Ferrosilicon Futures (Silicon Ferroalloy) |

| August 8, 2014 | Zhengzhou Commodity Exchange | Manganese silicon futures (manganese silicon alloy) |

At present, the size of China's commodity futures market continues to expand, and it has initially possessed the market capacity required for large-scale asset allocation. China's commodity futures ranked first in the world's trading volume for nine consecutive years from 2010 to 2018. Among them, the bilateral trading volume of China's commodity futures in 2016 was 8.238 billion contracts, of which 2.827 billion were in ferrous metal futures, accounting for 34.32% of China's commodity futures trading volume. From 2016 to 2018, the proportion of ferrous metal futures in China's commodity futures market has exceeded agricultural products, nonferrous metal futures and chemical futures that have been listed for a long time, becoming the largest trading volume in China's commodity futures market. The trading volume of commodity futures in China from 2014 to 2018 is shown in Table 2.

**Table 2: Comparison of bilateral trading volumes of China's commodity futures from 2014 to 2018 (unit of volume: million hands)[1]**

| Futures | From 2014 to 2018 | | | | |
|---|---|---|---|---|---|
| | 2018 | 2017 | 2016 | 2015 | 2014 |
| Ferrous Metal Futures | 2020.8 | 2514.82 | 2826.52 | 1668.16 | 1256.23 |
| Non-Ferrous Metal Futures | 752.71 | 744.62 | 838.19 | 783.95 | 687.35 |
| Agricultural Futures | 1713.49 | 1447.28 | 2679.3 | 2151.82 | 1738.45 |
| Chemical Futures | 1479.98 | 1386.21 | 1894.97 | 1870.37 | 894.51 |
| Total | 5966.98 | 6092.93 | 8238.98 | 6474.3 | 4576.54 |

Not so many scholars pay attention to the spillover effect on the volatility of ferrous metal futures due to the late listing time of ferrous metal futures. However, research on the spillover effect of agricultural futures and nonferrous metal futures is of great significance. Among assets, commodities serve as diversifiers in the process of portfolio choice. Volatility is an important metric indicating uncertainty or risk. Asset return comovements and transmission of volatility shocks have significant implications for asset pricing and portfolio allocation (Aloui et al., 2012). Our study focuses on volatility spillovers associated with financial markets, defined as the intensity of price fluctuations in financial instruments (Seth & Sing Hania, 2018).

**3.2 Data selection and descriptive statistics**

To study the volatility spillover effects between ferrous metal futures and other sectors of Chinese commodity futures, we use the commodity futures index as a reference. We select the ferrous metal commodity futures index(JJRI.WI), precious metal index(NMFI.WI), nonferrous metal index(NFFI.WI), energy index(ENFI.WI) and the Chemical Index(CIFI.WI) are respectively used as the measurement index of ferrous metal futures, precious metal futures, nonferrous metal futures, energy futures and chemical futures. We analyze the volatility spillover effects between the above commodity futures sectors. The futures sector classification and specific futures varieties included are shown in Table 3. The classification and specific data of the above indices are derived from the Chinese Wind Database.

---
[1]Data source: refer to the classification in the Wind Commodity Index statistics

Table 3: Futures sector index and its classification[2]

| Futures Sector Index | Specific futures varieties |
|---|---|
| ferrous metal commodity futures Index(JJRI.WI) | rebar, hot-rolled coil, wire, iron ore, coking coal, coke, ferrosilicon, manganese, silicon |
| Precious metal index (NMFI.WI) | gold, silver |
| nonferrous metal index(NFFI.WI) | copper, aluminum, lead, zinc, nickel, tin |
| energy index(ENFI.WI) | Fuel oil, thermal coal, crude oil |
| Chemical Index(CIFI.WI) | LLDPE, polypropylene, PTA, methanol, rubber, asphalt |

We select data from the Wind database for 2877 trading days from March 30, 2009, to January 21, 2021. DNLHSJS, DLNGJS, DLNYSJS, DLNNY, and DNLHG, respectively, stand for the ferrous metal commodity futures index, precious metal index, nonferrous metal index, energy index, and chemical index. Table 4 is the descriptive statistics of the data.

Table 4: Descriptive statistics of the logarithmic return series of the futures sector index

| Logarithmic returns | | DLNHSJS | DLNGJS | DLNYSJS | DLNNY | DLNHG |
|---|---|---|---|---|---|---|
| mean | | 0.0000 | 0.0001 | 0.0002 | -0.0001 | -0.0001 |
| Maximum value | | 0.0659 | 0.0656 | 0.0562 | 0.1128 | 0.0612 |
| Minimum value | | -0.0759 | -0.0930 | -0.0627 | -0.1779 | -0.0715 |
| Standard deviation | | 0.0142 | 0.0123 | 0.0125 | 0.0140 | 0.0147 |
| Skewness | | -0.0028 | -0.3146 | -0.2142 | -0.6532 | -0.1929 |
| Kurtosis | | 6.1062 | 7.8162 | 6.0955 | 18.5309 | 4.5170 |
| Number of samples | | 2877 | 2877 | 2877 | 2877 | 2877 |
| normality test | JB | 1156.61 | 2828.07 | 1170.66 | 29119.48 | 293.70 |
| | P value | 0.00 | 0.00 | 0.00 | 0.00 | 0.00 |
| Stationarity test of ADF | t | -53.75 | -54.35 | -56.85 | -34.65 | -53.88 |
| | P | 0.00 | 0.00 | 0.00 | 0.00 | 0.00 |

From Table 4, it can be concluded that the Jarque-Bera statistics of the logarithmic return series of each commodity futures sector index significantly exceed the threshold of hypothesis testing that rejects the normal distribution, so it does not recognize that the sample probability follows the normal distribution. It can be seen from the stationary test of ADF that the null hypothesis is rejected, in which unit roots exist. The logarithmic return series of each commodity futures sector index are all stable time series.

---
[2] Data Source: Chinese Wind database

**3.3 Dynamic spillover effect under the DCC-GARCH model**

The GARCH (1, 1) is established for the logarithmic return sequence of every futures index to obtain the dynamic correlation coefficient of the volatility between ferrous metal futures and other plates in Chinese commodity futures. The parameters of the GARCH model are estimated, and the results can be shown in Table 5. The coefficient of the ARCH (1) is α, and the coefficient of the GARCH (1) is β. Based on the estimation results of the GARCH (1, 1) model, the maximum likelihood estimation is performed based on the DCC equation. $q_{ij,t}$ is the conditional variance of the dynamic correlation coefficient between ferrous metal futures and commodity futures in other sectors. The estimated results of the parameters θ and η in the DCC equation coefficients are listed in Table 5.

**Table 5: Estimated results based on DCC-GARCH (1,1) model**

| sequence | coefficient $\alpha$ for ARCH(1) | coefficient $\beta$ for GARCH(1) | DCC equation |
|---|---|---|---|
| Ferrous Metals Futures | 0.0622***(7.6635) | 0.9409***(129.0421) | $\theta$ |
| Precious Metals Futures | 0.0407***(7.8402) | 0.9587***(198.7621) | 0.0168***(8.8242) |
| Nonferrous metal futures | 0.0530***(8.5161) | 0.9401***(138.6406) | $\eta$ |
| Energy futures | 0.0651***(7.3473) | 0.9300***(108.3730) | 0.9786***(364.6919) |
| Chemical futures | 0.0527***(8.0687) | 0.9443***(151.2474) | maximum likelihood 45725.5 |
| Note:"***"、"**"、"*" represents the significance level of 1%, 5% and 10% respectively. Moreover, the corresponding t values are in the parentheses. | | | |

It can be seen from Table 5 that, at the significance level of 1%, the ARCH (1) effect and GARCH (1) effect of various commodity futures sectors are widespread. Moreover, the sum of the coefficients for α and β of each variety is close to 1, indicating that the volatility of each commodity futures sector has the characteristics of aggregation and persistence. θ and η, the parameters of the correlation coefficient between various futures sector indexes, pass the test at a significance level of 1%, indicating that the dynamic correlation between ferrous metal futures and other sector commodity futures is obvious.

**Table 6: Mean values of dynamic correlation coefficients between ferrous metals and other commodity futures under the DCC-GARCH model**

| Correlation coefficient | E(ρ) |
|---|---|
| between ferrous and precious metals | 0.156***(70.5660) |

| between ferrous and nonferrous metals | 0.455***(226.4885) |
|---|---|
| between ferrous metals and energy futures | 0.331***(157.9518) |
| between ferrous metals and chemical futures | 0.505***(306.1529) |
| Note:"***"、"**"、"*"represents the significance levels of 1%, 5%, and 10%, respectively, and the corresponding Z values are in parentheses; | |

At a significance level of 1%, the average dynamic correlation coefficient between ferrous metal futures and chemical industry futures is as high as 0.505. There is a strong correlation between them. At a significance level of 1%, the average dynamic correlation coefficient between ferrous and nonferrous metal futures is 0.455. The average value of the dynamic correlation coefficient between ferrous metals and energy futures is 0.331. The average value of the dynamic correlation coefficient between ferrous and precious metal futures is the lowest, only 0.156.

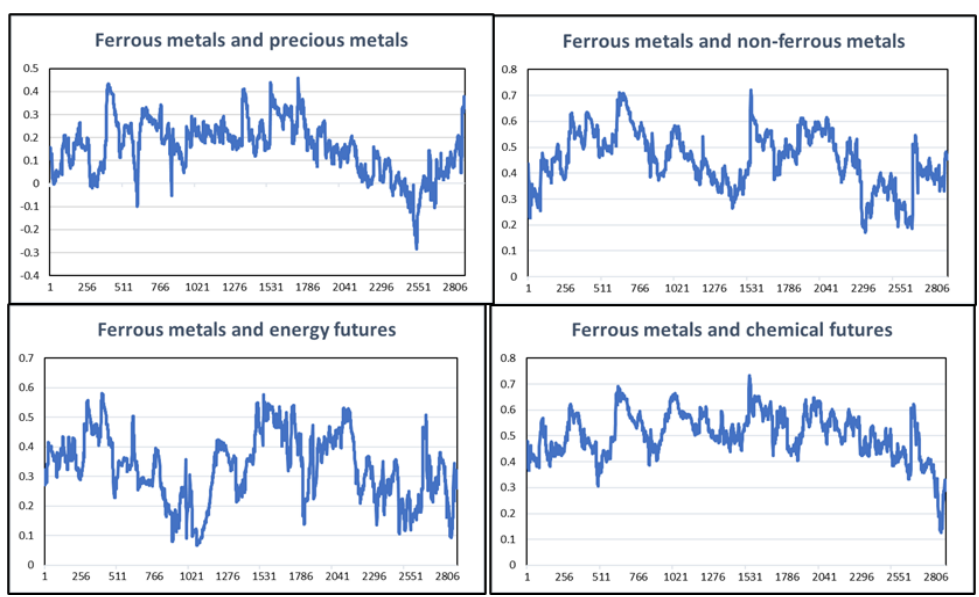

**Figure 2: Dynamic correlation coefficient between ferrous metal futures and other commodity futures under the DCC-GARCH model**

As seen from Table 6 and Figure 2, at a significance level of 1%, the volatility spillover effects between ferrous metal and nonferrous metal futures, ferrous metal and energy futures, and ferrous metal and chemical futures are obvious. However, the transmission direction of the spillover effect cannot be determined by the dynamic correlation coefficient, which requires constructing a BEKK-GARCH model.

### 3.4 Volatility spillover effects under the BEKK-GARCH model

**Table 7: BEKK-GARCH (1,1) model estimation results**

| variable | parameter | variable | parameter |
|---|---|---|---|
| A(1,1) | 0.1844***(12.2650) | B(1,1) | 0.9834***(430.4747) |
| A(1,2) | -0.0018(-0.2152) | B(1,2) | -0.0012(-0.8303) |
| A(1,3) | -0.0012(-0.1023) | B(1,3) | 0.0006(0.2763) |
| A(1,4) | -0.0532***(-3.9158) | B(1,4) | 0.0107***(4.1241) |
| A(1,5) | -0.0299**(-2.1316) | B(1,5) | 0.0069***(2.6884) |
| A(2,1) | -0.0118(-1.2601) | B(2,1) | 0.0044***(2.7138) |
| A(2,2) | 0.1749***(13.7508) | B(2,2) | 0.9821***(437.0265) |
| A(3,1) | 0.0263**(2.1036) | B(3,1) | -0.0125***(-4.1438) |
| A(3,3) | 0.1955***(12.6282) | B(3,3) | 0.9705***(226.8011) |
| A(4,1) | -0.0032(-0.2711) | B(4,1) | 0.0027(1.0194) |
| A(4,4) | 0.2061***(13.6814) | B(4,4) | 0.9744***(302.2258) |
| A(5,1) | -0.0073(-0.7408) | B(5,1) | 0.0039**(2.5145) |
| A(5,5) | 0.1848***(14.4232) | B(5,5) | 0.9739***(208.2013) |

Note: "***"、"**"、"*" represents the significance level of 1%, 5%, and 10%, respectively and the corresponding t values are in the parentheses.

First, define $i = 1, 2, 3, 4, 5; j = 1, 2, 3, 4, 5$. Moreover, $1, 2, 3, 4$ and $5$ represent the logarithmic return sequence of ferrous, precious, nonferrous, energy, and chemical futures.

The matrix elements A(1,2), B(1,2), A(2,1) of the asymmetric term coefficient are not significant at the significance level of 10%, while B(2,1) are significant at a significance level of 1%. These indicate that precious metal futures have a volatility spillover effect on ferrous metal futures and can not be reversed. The GARCH effect only reflects it, and the ARCH effect is not significant. There is no volatility spillover effect of ferrous metal futures on precious metal futures.

The coefficients of A(1,3) and B(1,3) are not significant at a significance level of 10%. While A(3,1) coefficients are significant at a significance level of 5%, B(3,1) is significant at a significance level of 1%. It shows a volatility spillover effect of nonferrous metal futures on ferrous metal futures.

The coefficients of A(1,4) and B(1,4) are significant at the significance level of 1%, while the coefficients of A(4,1) and B(4,1) are not significant at the significance level of 10%. Significantly, it indicates a one-way fluctuation spillover effect of ferrous metal futures on energy futures, while the fluctuation overflow effect of energy futures on ferrous metal futures is insignificant.

The coefficients of A(1,5) and B(5,1) are significant at the significance level of 5%, B(1,5) is significant at the significance level of 1%. At the same time, the coefficients of A(5,1) are not obvious at the significance level of 10%. It shows the fluctuation spillover effect of ferrous metal futures on chemical futures. Although the chemical futures' volatility spillover effect on ferrous metal futures also exists, it is only reflected by the GARCH effect, and the ARCH effect is not significant.

Based on the BEKK-GARCH model, the relationship and direction of volatility spillover effects between ferrous metal futures and other sector commodity futures are shown in Figure 3.

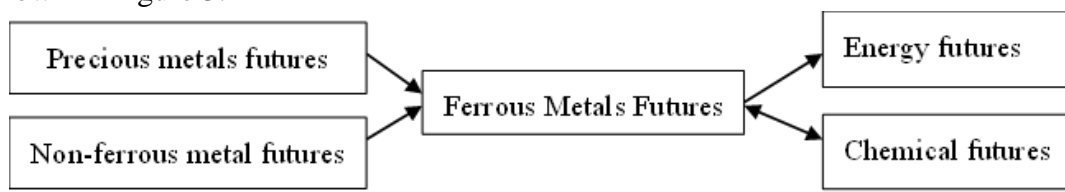

**Figure 3: Volatility spillover effects between ferrous metal futures and other sector commodity futures**

From Figure 3, we can see the conduction direction of the volatility spillover effect between ferrous metal futures and commodity futures in precious metals, nonferrous metals, energy, and chemical sectors. There exists a mutual volatility spillover relationship between ferrous metal futures and chemical futures. There only exists a volatility spillover effect of precious metals futures on ferrous metals futures, nonferrous metal futures on ferrous metal futures, and ferrous metal futures on energy futures.

**3.5** Directional spillovers under the DY(2012) index

We use DY index in a substantive empirical analysis of daily volatility spillovers across ferrous metal futures, precious metal futures, nonferrous metal futures, energy futures, and chemical futures over a daily time from March 1, 2017, to January 21, 2021, for a total of 952 daily observations, including the recent COVID-19 virus situation.

We use high and low prices daily to calculate the annualized daily percent standard deviation (volatility). Then we plot the five indexes' volatility in Figure 4.

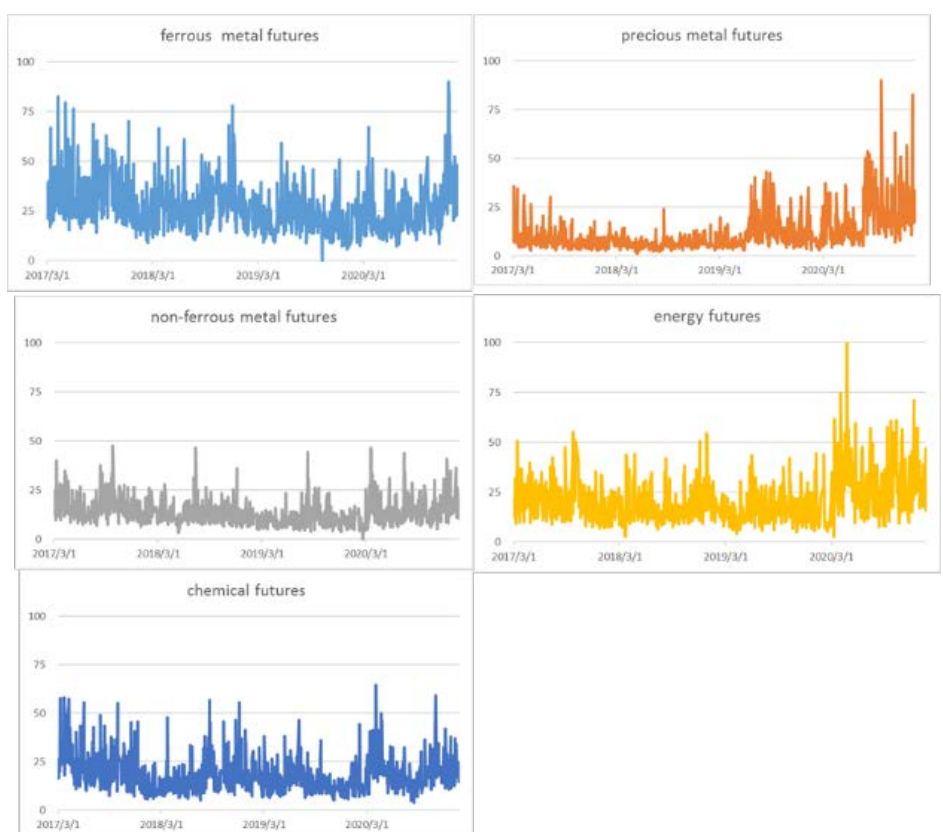

**Figure 4: Daily volatility of five commodity futures**

From figure 4 above, we can find that: (1) The ferrous metal futures have been the most volatile. (2) All volatilities are high during the recent COVID-19 virus situation in 2020.

**Total Spillovers**

The total spillover index measures the contribution of spillovers of volatility shock across five indexes to the total forecast error variance.

We call Table 8 a volatility spillover table. Its ij$^{th}$ entry is the estimated contribution to the forecast error variance of market i coming from innovation to market j. Hence, the off-diagonal column sums (labeled directional to others) or row sums (labeled directional from others), are "to" and "from" directional spillovers. In addition, the total volatility spillover index appears in the lower right corner of the spillover table. It is approximately the grand off-diagonal column sum (or row sum) relative to the grand column sum, including diagonals (or row sum including diagonals), expressed as a percent. The volatility spillover table provides an approximate "input-output" decomposition of the total volatility spillover index.

**Table 8: Volatility Spillover Table for commodity futures indexes**

| FROM / TO | Ferrous Metals Futures | Precious Metals Futures | Nonferrous metal futures | Energy futures | Chemical futures | Directional From Others |
|---|---|---|---|---|---|---|
| Ferrous Metals Futures | 69.49 | 1.50 | 10.35 | 5.64 | 13.02 | 30.51 |
| Precious Metals Futures | 1.93 | 88.04 | 3.11 | 5.11 | 1.80 | 11.96 |
| Nonferrous metal futures | 8.34 | 4.95 | 70.50 | 5.01 | 11.20 | 29.50 |
| Energy futures | 5.12 | 4.51 | 7.02 | 70.19 | 13.16 | 29.81 |
| Chemical futures | 9.70 | 2.18 | 10.21 | 9.26 | 68.65 | 31.35 |
| Directional TO Others | 25.10 | 13.13 | 30.69 | 25.02 | 39.18 | Total Spillover Index=(133.13/500) |
| Directional Including Own | 94.60 | 101.17 | 101.19 | 95.21 | 107.83 | **26.63%** |

From the "directional to others" row, we see that the gross directional volatility spillovers to others from each of the five indexes are very different. Ferrous metal futures have an evident spillover on others of about 25 percent. We also see from the "directional from others" column that gross directional volatility spillovers for ferrous metal futures are relatively large, at 30.51 percent.

Consider the total (non-directional) volatility spillover, which effectively distills the various directional volatility spillovers into a single index. The total volatility spillover is 26.63%, which indicates that, on average, across our entire sample, 26.63 percent of the volatility forecast error variance in all five indexes comes from spillovers. The summary of Table 8 is simple: Total and directional spillovers over the full sample period were evident.

The full-sample spillover table and spillover index constructed earlier, although providing a useful summary of "average" volatility spillover behavior, likely miss potentially important secular and cyclical movements in spillovers. To address this issue, we now estimate the volatility spillovers using 104-day rolling samples and a forecast horizon of 10-day, and we assess the extent and the nature of spillover variation overtime via the corresponding time series of spillover indices, which we examine graphically in the so-called total spillover plot of Figure 5.

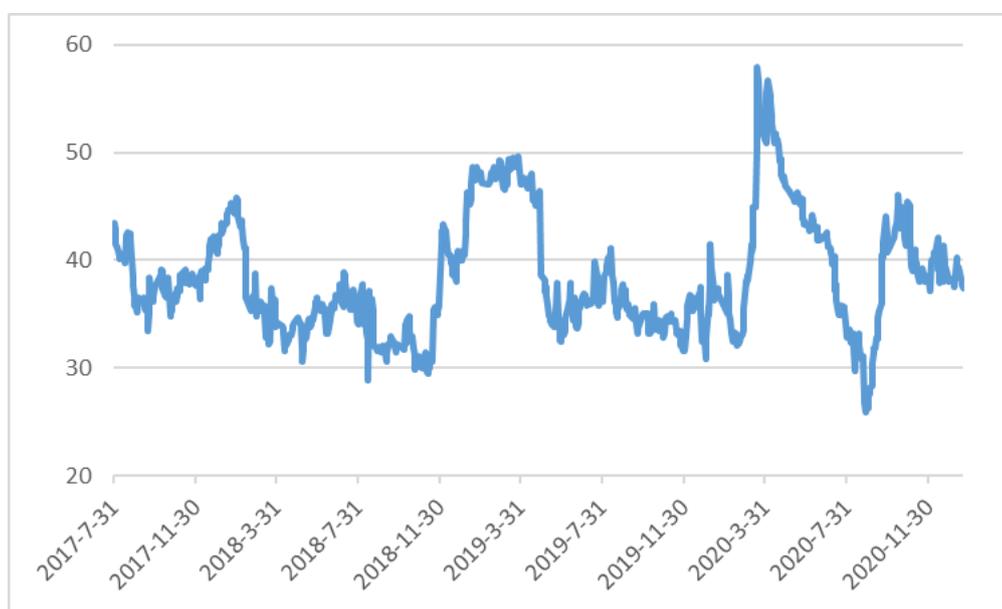

**Figure 5: Total Volatility Spillovers, Five indexes**

Starting at a value of 43.32 percent in the first window, the total volatility spillover plot for most of the time fluctuates between 25 and 50 percent. However, there are important exceptions: The spillovers exceeded the 50 percent mark in March 2020 and reached a 60 percent level at the beginning of the covid-19 virus situation.

**Directional Spillovers**

In Figure 6, we present the dynamic directional volatility spillovers from ferrous metal futures to others (corresponding to the "directional to others" row in Table 8) using 104-day rolling samples and a forecast horizon of 10-days. They vary greatly from 15 percent to 60 percent over time. The spillovers exceeded the 50 percent mark in March 2020 at the beginning of the covid-19 virus situation, but this is not its max value in history.

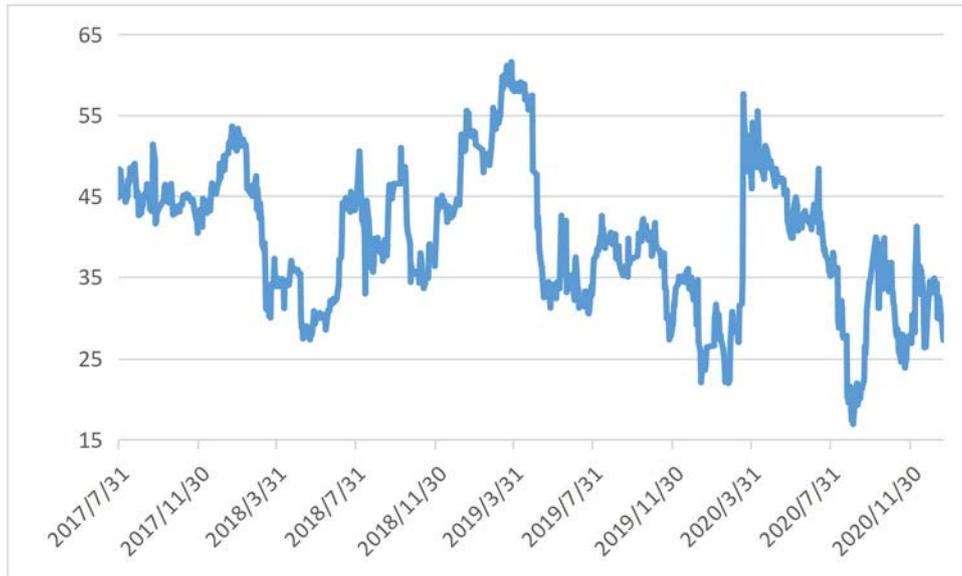

**Figure 6: Directional Volatility Spillovers, Ferrous metal futures with Four indexes**

In Figure 7, we present directional volatility spillovers from others to Ferrous metal futures (corresponds to the "directional from others" column in Table 2). The spillovers from others to Ferrous metal futures also vary noticeably over time. However, it fluctuated at the bottom in 2020 with the covid-19 virus situation. The relative variation pattern is reversed.

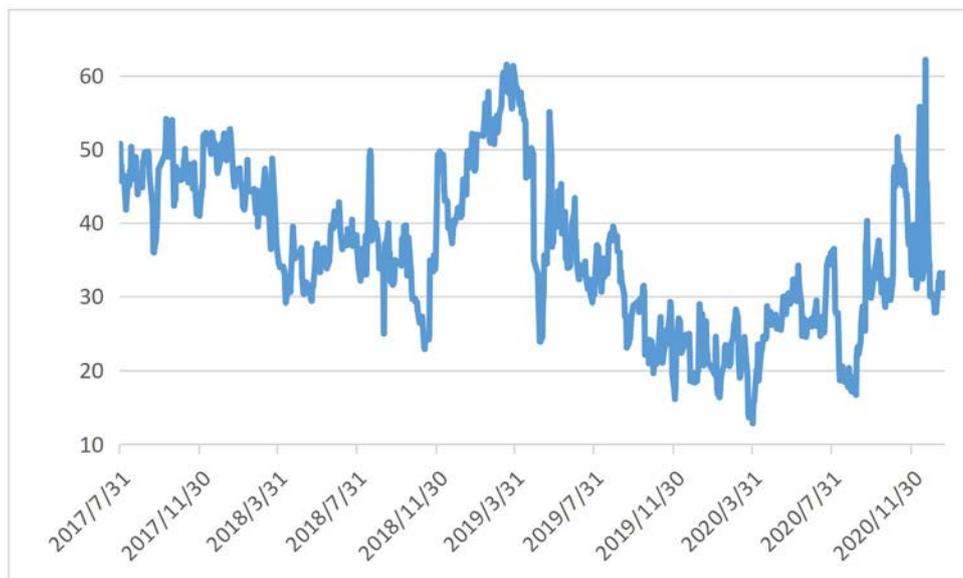

**Figure 7: Directional Volatility Spillovers, Ferrous metal futures with Four indexes Net Spillovers**

We obtain the rolling-sample net volatility spillover from ferrous metal futures to all other indexes. The net volatility spillover is simply the difference between the gross volatility shock transmitted to and the gross volatility shocks received from all other indexes, presented in Figure 8.

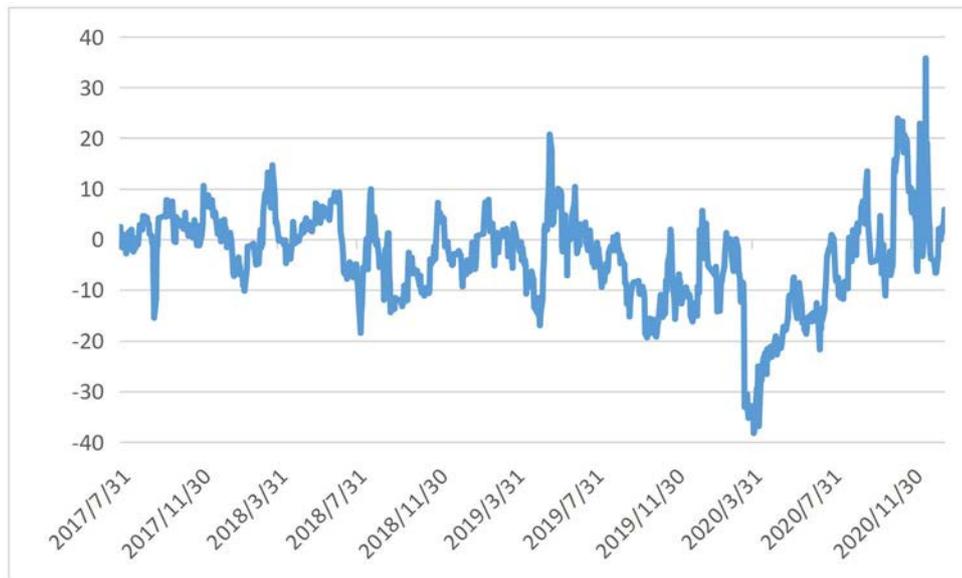

**Figure 8: Net Volatility Spillovers, Ferrous metal futures to all other indexes**

From Figure 8, the net Volatility Spillovers of ferrous metal futures fluctuate between -40 percent and 40 percent, which shows that ferrous metal futures can be spillover transmitters or recipients. In 2020, with the covid-19 virus situation, it was a spillover recipient at the beginning, then became a spillover transmitter at the end of the year.

**Net Pairwise Spillovers**

We also calculate rolling-sample net pairwise spillovers between the two indexes and present these plots in Figure 9. The net pairwise volatility spillover between markets i and j is the difference between gross volatility shock transmitted from market i to j and gross volatility shocks transmitted from j to i.

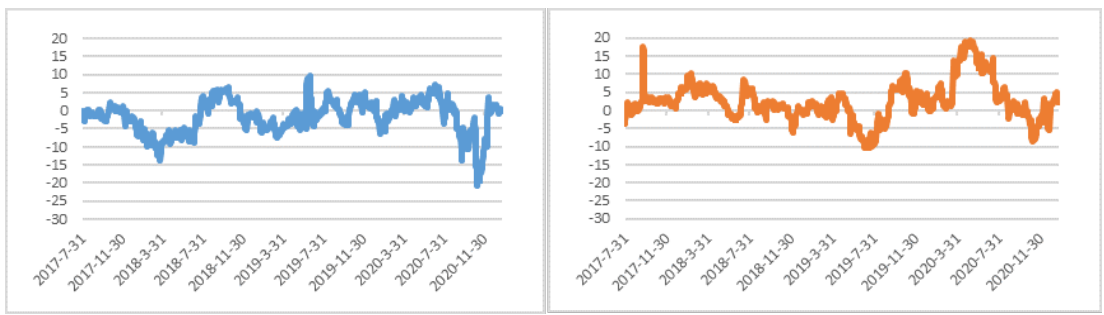

a) ferrous metal futures–precious metal futures    b) ferrous metal futures–nonferrous metal futures

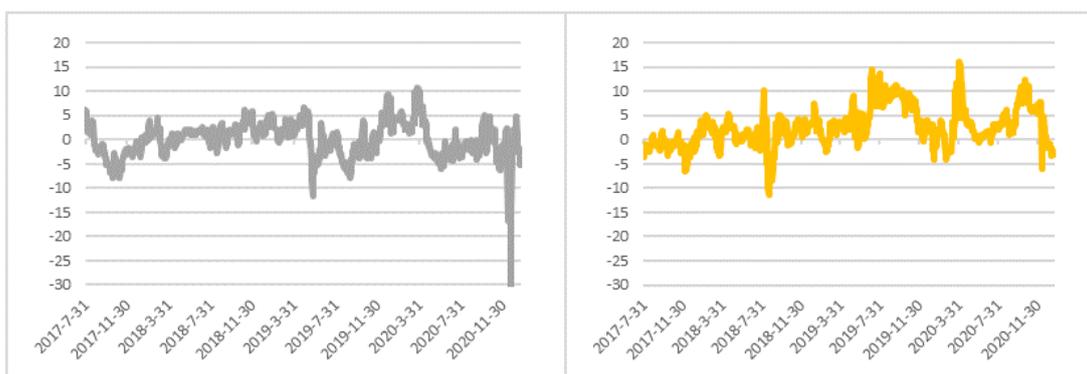

c) ferrous metal futures-energy futures    d) ferrous metal futures-chemical futures

Figure 9: Net Pairwise Volatility Spillovers, ferrous metal futures to other indexes

From Figure 9, we can find that: 1) ferrous metal futures are the main spillover recipient from precious metal futures and energy futures, ferrous metal futures to energy futures reach the min value of -30 percent in December 2020, receiving the max spillover from energy futures. 2) ferrous metal futures are meanly a spillover transmitter to nonferrous metal futures and chemical futures. Ferrous metal futures to nonferrous metal futures reached the max value of 18 percent in May 2020.

**4 Spillover relationship between ferrous metal futures and Chinese stock market**

**4.1 Descriptive statistics**

To examine the impact of ferrous metal futures on the stock market, we select the stock price indexes of the steel industry, real estate industry, building materials industry, automobile manufacturing industry, machinery and equipment industry, and home appliance industry in the Chinese stock market as the indicators of the corresponding industry sectors. All stock price indexes are from the Shenwan first-level industry classification issued by SWS Research3. We select data from the Wind database for a total of 2877 trading days from March 30, 2009, to January 21, 2021. We have named DLNHJQH, DLNGTZS, DLNJYDQZS DLNFDCZS, DLNJZCLZS, DLNQCZS, and DLNJXSBZS stand for the yield sequence of the ferrous metal futures, the steel industry stock price index, the home appliance industry index, the real estate industry stock price index, the building materials industry stock price index, the automobile manufacturing industry stock price index, and the mechanical equipment industry stock price index, the descriptive statistics of the data can be shown in table 9.

**Table 9: Descriptive statistics of yield series of the logarithmic price index**

| yield series of the logarithmic price index | DLNHJQH | DLNGTZS | DLNJYDQZS | DLNFDCZS | DLNJZCLZS | DLNQCZS | DLNJXSBZS |
|---|---|---|---|---|---|---|---|

---

[3] Full name: Shanghai Shenyin Wanguo Securities Research Institute Co., Ltd.

| | | | | | | | |
|---|---|---|---|---|---|---|---|
| Mean | | 0.0000 | 0.0000 | 0.0007 | 0.0001 | 0.0004 | 0.0004 | 0.0003 |
| Maximum value | | 0.0659 | 0.0743 | 0.0732 | 0.0745 | 0.0813 | 0.0717 | 0.0696 |
| Minimum value | | −0.0759 | −0.1030 | −0.0923 | −0.0973 | −0.0917 | −0.0969 | −0.1000 |
| Standard deviation | | 0.0142 | 0.0182 | 0.0175 | 0.0184 | 0.0188 | 0.0178 | 0.0177 |
| Skewness | | −0.0028 | −0.5463 | −0.3937 | −0.5898 | −0.5929 | −0.6394 | −0.8481 |
| Kurtosis | | 6.1062 | 6.2694 | 5.4037 | 6.2374 | 5.7191 | 6.2032 | 6.4700 |
| Number of samples | | 2877 | 2877 | 2877 | 2877 | 2877 | 2877 | 2877 |
| normality test | JB | 1156.61 | 1424.47 | 766.94 | 1423.20 | 1054.83 | 1426.00 | 1788.37 |
| | P Value | 0.00 | 0.00 | 0.00 | 0.00 | 0.00 | 0.00 | 0.00 |
| Stationarity test of ADF | ADF | −53.75 | −52.13 | −52.29 | −51.55 | −49.91 | −50.56 | −49.77 |
| | P Value | 0.00 | 0.00 | 0.00 | 0.00 | 0.00 | 0.00 | 0.00 |

From the stationarity test of ADF, it can be seen that the yield series of the logarithmic price index are stationary.

**4.2 Research on dynamic correlation under the DCC-GARCH model**

First, the GARCH (1,1) model is established for each logarithmic return sequence of the price index, and the parameters are estimated. The parameters of the GARCH model are estimated, and the results can be shown in Table 10. The coefficient of the ARCH (1) is α, and the coefficient of the GARCH (1) is β. Based on the estimation results of the GARCH (1, 1) model, the maximum likelihood estimation is performed based on the DCC equation. $q_{ij,t}$ is the conditional variance of the dynamic correlation coefficient between ferrous metal futures and the stock price indexes. The estimated results of the parameters θ and η in the DCC equation coefficients are in Table 10.

**Table 10: Estimated results of ferrous metals futures and stock industry indexes based on the DCC-GARCH model**

| Sequence | Coefficient α for ARCH(1) | Coefficient β for GARCH(1) | DCC equation |
|---|---|---|---|
| DLNHJQH | 0.0705***(8.0796) | 0.9312***(115.3341) | θ |
| DLNGTZS | 0.0507***(8.1780) | 0.9338***(120.0354) | 0.0210***(13.4716) |
| DLNJYDQZS | 0.0452***(8.2709) | 0.9382***(122.4515) | η |
| DLNFDCZS | 0.0501***(9.6323) | 0.9394***(152.0912) | 0.9708***(382.8091) |
| DLNJZCLZS | 0.0437***(11.1527) | 0.9447***(191.7510) | Maximum Likelihood |
| DLNQCZS | 0.0438***(10.7189) | 0.9470***(187.1368) | |
| DLNJXSBZS | 0.0421***(10.5736) | 0.9476***(194.3362) | 64891.7081 |
| Note:"***"、"**"、"*" represents the significance level of 1%, 5%, and 10%, respectively, and the corresponding t values are in the parentheses. | | | |

It can be seen from Table 9 that at the significance level of 1%, the ARCH (1) effect and GARCH (1) effect are widespread. And the sum of the coefficients for α and β GARCH(1) of each vsariety is close to 1, indicating that the volatility of each sector has the characteristics of aggregation and persistence. θ and η, which are the parameters of the correlation coefficient, pass the test at a significance level of 1%, indicating an obvious dynamic correlation between ferrous metal futures and the stock price indexes.

**Table 11: Mean values of dynamic correlation coefficients between ferrous metals and stock price indexes under the DCC-GARCH model**

| Correlation coefficient | E($\rho$) |
|---|---|
| between ferrous and steel industry price index | 0.3494***(173.0828) |
| between ferrous and home appliance industry price index | 0.2264***(94.7068) |
| between ferrous and real estate industry price index | 0.2159***(89.3861) |
| between ferrous and building materials industry price index | 0.2649***(117.6260) |
| between ferrous and automobile manufacturing industry price index | 0.2382***(103.6987) |
| between ferrous and mechanical equipment industry price index | 0.2429***(104.0497) |

At a significance level of 1%, the average dynamic correlation coefficient between ferrous metal futures and steel industry stock price index is 0.349, the average dynamic correlation coefficient between ferrous metal futures and home appliance industry price index is 0.226, the average dynamic correlation coefficient between ferrous metal futures and real estate industry price index is 0.216, the average dynamic correlation coefficient between ferrous metal futures and building materials industry price index is 0.265, the average dynamic correlation coefficient between ferrous metal futures and automobile manufacturing industry price index is 0.238, the average dynamic correlation coefficient between ferrous metal futures and mechanical equipment industry price index is 0.243.

### 4.3 Volatility spillover effects under the BEKK-GARCH model

**Table 12: BEKK-GARCH (1,1) model estimation results**

| variable | parameter | variable | parameter |
|---|---|---|---|
| A(1,1) | 0.1867***(13.1023) | B(1,1) | 0.9812***(367.5036) |
| A(1,2) | -0.0036(-0.4050) | B(1,2) | 0.0018(1.0819) |
| A(1,3) | 0.0279***(3.7925) | B(1,3) | -0.0074***(-5.3758) |
| A(1,4) | 0.0109(1.5031) | B(1,4) | -0.0018(-1.3979) |
| A(1,5) | -0.0010(-0.1338) | B(1,5) | 0.0001(0.0783) |
| A(1,6) | 0.0045(0.7076) | B(1,6) | -0.0008(-0.7110) |
| A(1,7) | -0.0030(-0.4302) | B(1,7) | 0.0002(0.1781) |
| A(2,1) | 0.0241*(1.9229) | B(2,1) | -0.0026(-0.7973) |
| A(2,2) | 0.2428***(16.4102) | B(2,2) | 0.9630***(230.2849) |
| A(3,1) | 0.0077(0.6111) | B(3,1) | 0.0044*(1.9521) |
| A(3,3) | 0.1161***(7.7059) | B(3,3) | 0.9953***(390.4249) |
| A(4,1) | -0.0175(-1.5265) | B(4,1) | 0.0020(0.7476) |
| A(4,4) | 0.2286***(16.2916) | B(4,4) | 0.9652***(306.8873) |
| A(5,1) | -0.0037(-0.2830) | B(5,1) | -0.0010(-0.4754) |

| variable | parameter | variable | parameter |
|---|---|---|---|
| A(5,5) | 0.1740***(12.4390) | B(5,5) | 0.9819***(322.9530) |
| A(6,1) | 0.0175(1.2742) | B(6,1) | -0.0025(-1.0863) |
| A(6,6) | 0.2095***(11.2572) | B(6,6) | 0.9710***(294.8646) |
| A(7,1) | -0.0183(-1.2545) | B(7,1) | -0.0026(-0.9922) |
| A(7,7) | 0.0138(0.7751) | B(7,7) | 1.0047***(491.3894) |
| Note:"***"、"**"、"*" represents the significance level of 1%, 5%, and 10%, and the corresponding t values are in the parentheses. ||||

First, define i=1, 2, 3, 4, 5, 6, 7; j=1, 2, 3, 4, 5, 6, 7. Moreover, 1, 2, 3, 4, 5, 6 and 7 represent the logarithmic return sequence of ferrous metal futures, the steel industry stock price index, the home appliance industry index, the real estate industry stock price index, the building materials industry stock price index, the automobile manufacturing industry stock price index, and the mechanical equipment industry stock price index.

The matrix elements A(1,2),B(1,2),B(2,1) of the asymmetric term coefficient are not significant at a significance level of 10%, while A(2,1) is significant at a significance level of 10%. These indicate a volatility spillover effect from the steel stock price index on ferrous metal futures. Moreover, the volatility spillover effect of the steel stock price index on ferrous metal futures is only reflected through the ARCH effect.

The coefficients of A(1,3), B(1,3) are significant at a significance level of 1%, B(3,1) is significant at a significance level of 10%, A(3,1) is not significant at a significance level of 10%. It shows that the volatility spillover effect of ferrous metal futures on the home appliance industry index, and the reverse direction is only reflected by the GARCH effect.

The coefficients of A(1,4), B(1,4), A(4,1), B(4,1), A(1,5), B(1,5), A(5,1), B(5,1), A(1,6), B(1,6), A(6,1), B(6,1), A(1,7), B(1,7), A(7,1), B(7,1) are not significant at a significance level of 10%, it shows that there are no volatility spillover effect between ferrous metal futures and the real estate industry stock price index, between ferrous metal futures and the building materials industry stock price index, between ferrous metal futures and the automobile manufacturing industry stock price index, between ferrous metal futures and the mechanical equipment industry stock price index.

Above all, steel varieties are used as the raw material for manufacturing. The price of steel futures has a significant spillover effect on the price of products in the home appliance industry.

### 4.4 Directional spillovers under the DY(2012) index

We use DY index in a substantive empirical analysis of daily volatility spillovers across ferrous metal futures, the steel industry stock price index, the home appliance industry index, the real estate industry stock price index, the building materials industry stock price index, the automobile manufacturing industry stock price index, and the mechanical equipment industry stock price index over a daily period from March 1, 2017, to January 21, 2021, for a total of 952 daily observations, including the recent COVID-19 virus situation.

We use high and low prices daily to calculate the annualized daily percent standard deviation (volatility). We plot the seven indexes' volatility in Figure 10.

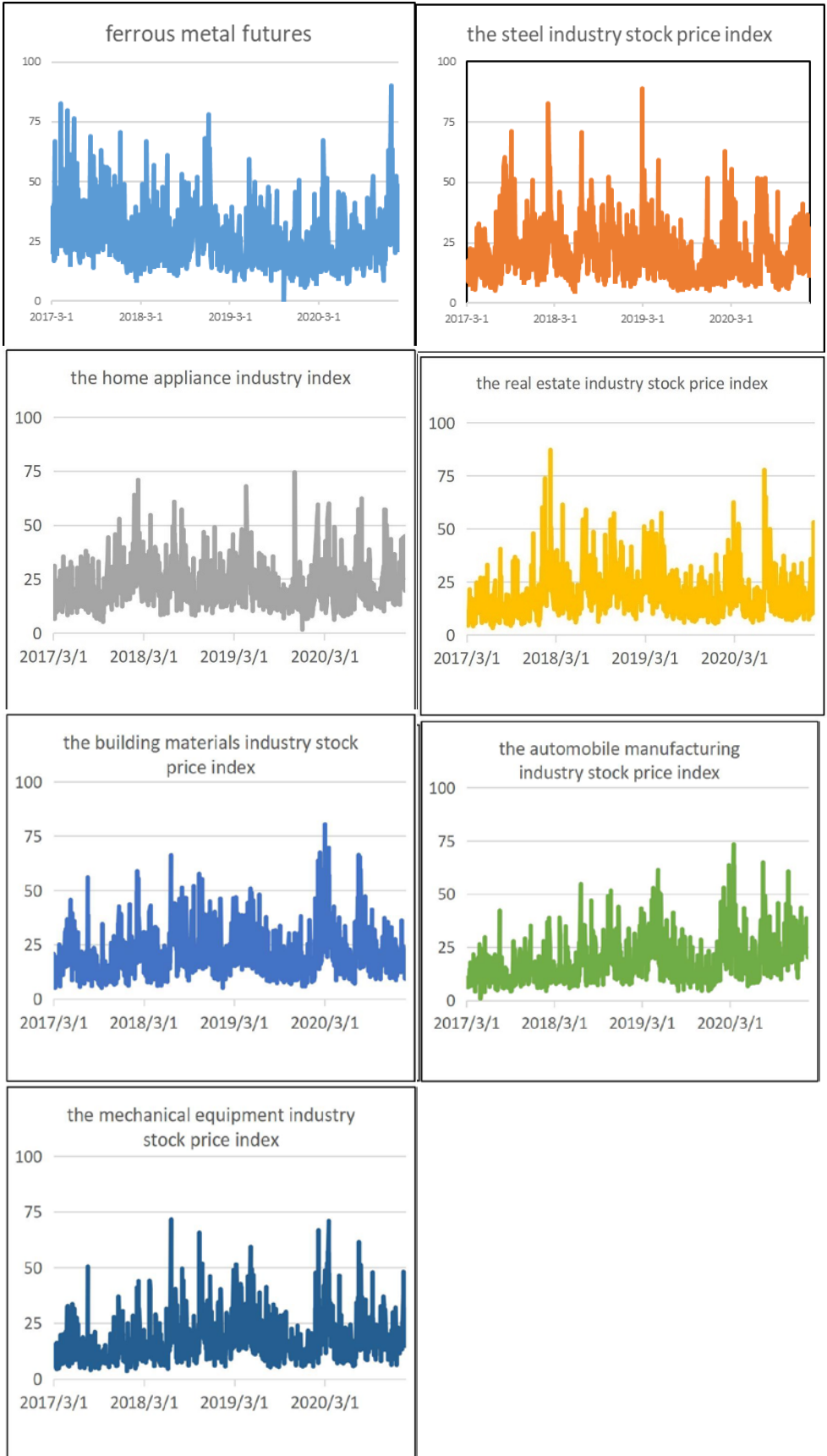

**Figure 10: Daily volatility of ferrous metal futures and six stock indexes**

From Figure 10, we can find that: (1) The ferrous metal futures have been the most volatile. (2) All volatilities except the steel industry stock price index were high during the recent COVID-19 virus in 2020.

**Total Spillovers**

The total spillover index measures the contribution of spillovers of volatility shocks across the seven indexes to the total forecast error variance.

We call Table 2 a volatility spillover table. Its $ij^{th}$ entry is the estimated contribution to the forecast error variance of market i coming from innovations to market j. The volatility spillover table provides an approximate "input-output" decomposition of the total volatility spillover index.

**Table 13: Volatility Spillover Table for stock price indexes**

| From \ To | 1. ferrous metal futures | 2. steel industry stock price index | 3. home appliance industry index | 4. real estate industry stock price index | 5. building materials industry stock price index | 6. automobile manufacturing industry stock price index | 7. mechanical equipment industry stock price index | Directional From Others |
|---|---|---|---|---|---|---|---|---|
| 1 | 94.22 | 2.29 | 0.22 | 0.56 | 0.66 | 1.17 | 0.88 | 5.78 |
| 2 | 0.68 | 46.86 | 4.12 | 10.27 | 15.72 | 9.72 | 12.63 | 53.14 |
| 3 | 0.26 | 5.30 | 46.32 | 11.51 | 10.52 | 14.06 | 12.03 | 53.68 |
| 4 | 0.56 | 8.96 | 7.23 | 39.56 | 16.19 | 11.58 | 15.93 | 60.44 |
| 5 | 0.36 | 11.32 | 6.35 | 13.41 | 33.68 | 14.75 | 20.13 | 66.32 |
| 6 | 0.51 | 6.98 | 8.48 | 8.67 | 13.00 | 38.39 | 23.97 | 61.61 |
| 7 | 0.53 | 9.00 | 6.25 | 11.04 | 17.43 | 22.22 | 33.53 | 66.47 |
| Directional TO Others | 2.90 | 43.85 | 32.64 | 55.47 | 73.52 | 73.50 | 85.56 | Spillover index=(367.44/700) |
| Directional Including Own | 97.11 | 90.71 | 78.96 | 95.03 | 107.20 | 111.89 | 119.10 | 52.49% |

From the "directional to others" row, we see that the gross directional volatility spillovers to others from each of the seven indexes are very different. Ferrous metal futures have a spillover effect to others of about 2.9 percent, which is smaller than the spillover of ferrous metal futures to other commodity futures, which is 25.1 percent. We also see from the "directional from others" column that the gross directional volatility spillovers from others to the ferrous metal futures is 5.78 percent, which is also smaller than the spillover of other commodity futures to ferrous metal futures which is 30.51 percent.

The total volatility spillover is 52.49%, which indicates that, on average, across our entire sample, 52.49 percent of the volatility forecast error variance in all seven indexes comes from spillovers. The summary of Table 13 is simple: Total and directional spillovers over the full sample period were evident. However, we can find both Directional From Others and Directional TO Others between the other six stock indexes are a bigger value than Ferrous metal futures with other indexes.

We now estimate the volatility spillovers using 104-day rolling samples and a forecast horizon of 10-day, and we assess the extent and the nature of spillover variation over time via the corresponding time series of spillover indices, which we examine graphically in the so-called total spillover plot of Figure 11.

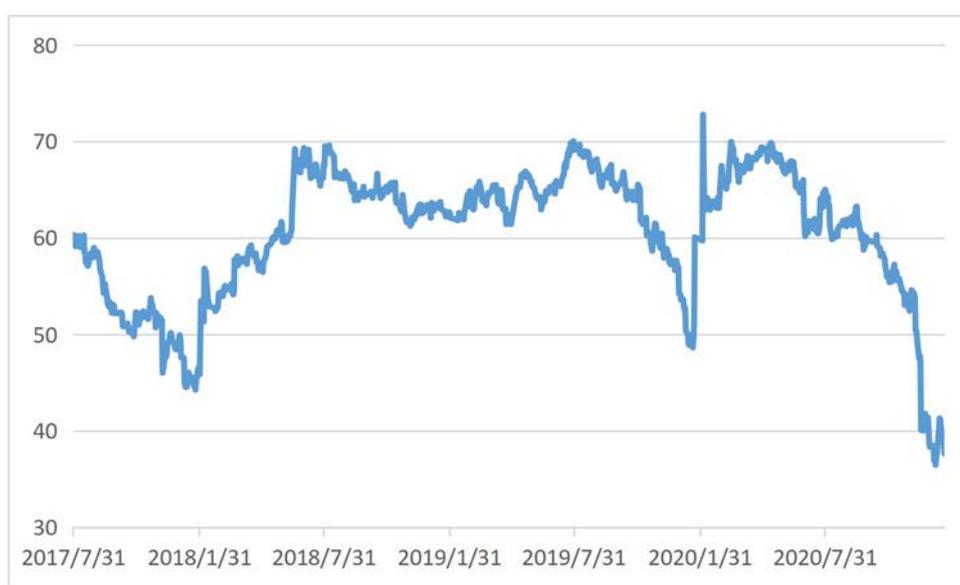

**Figure 11: Total Volatility Spillovers, ferrous metal futures and six stock indexes**

Starting at a value of 60 percent in the first window, the total volatility spillover plot for most of the time fluctuates between 45 and 70 percent. However, there are important exceptions: The spillovers reached a 72.76 percent level in February 2020, beginning the covid-19 virus situation. Then dropped down to a new low level of 36.47 in January 2021, after the covid-19 virus situation.

**Directional Spillovers**

In Figure 12, we present the dynamic directional volatility spillovers from ferrous metal futures to others (corresponding to the "directional to others" row in Table 13) using 104-day rolling samples and a forecast horizon of 10-days. They vary greatly from 10 percent to 70 percent over time. The spillovers reached a new high value of 69 percent in February 2020 at the beginning of the covid-19 virus situation and then dropped slowly during 2020.

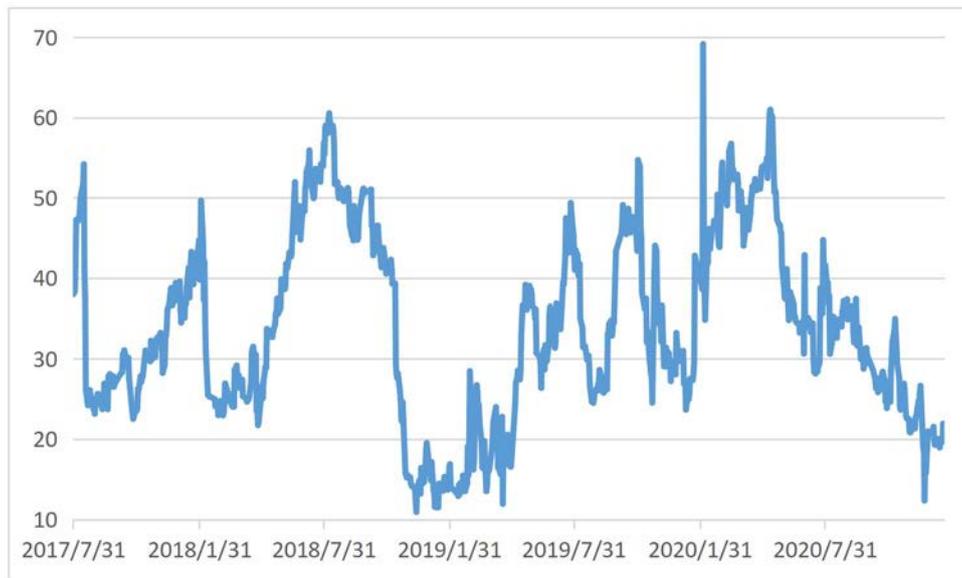

**Figure 12: Directional Volatility Spillovers, Ferrous metal futures to Six stock indexes**

In Figure 13, we present directional volatility spillovers from others to Ferrous metal futures (corresponding to the "directional from others" column in Table 13). The spillovers from others to Ferrous metal futures also vary noticeably over time. However, it fluctuated at the bottom in 2020 under the covid-19 virus situation, reaching a new high value in December 2020. The relative variation pattern is reversed.

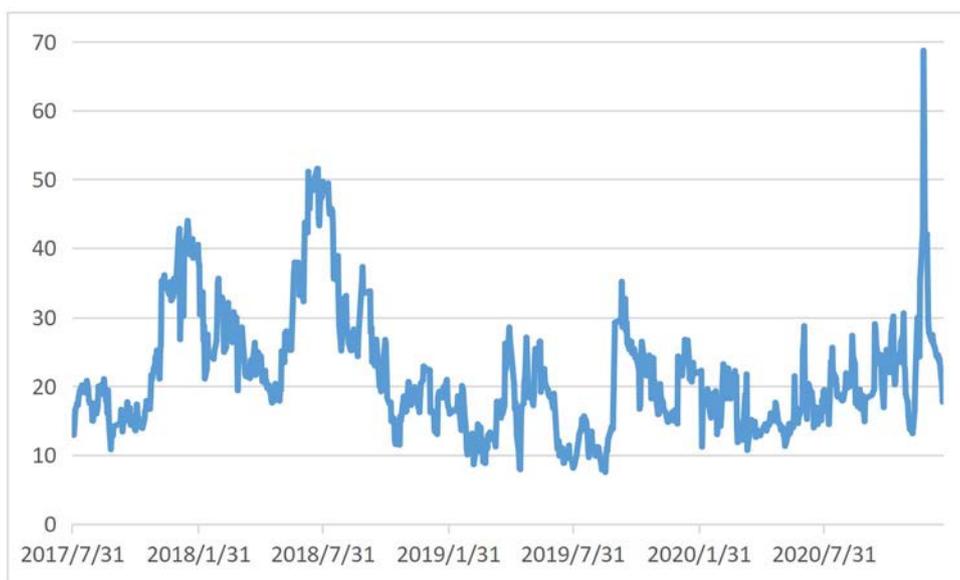

**Figure 13: Directional Volatility Spillovers, Ferrous metal futures from Six stock indexes**

**Net Spillovers**

We obtain the rolling-sample net volatility spillover from ferrous metal futures to all other markets j. The net volatility spillover is simply the difference between the gross volatility shocks transmitted to and the gross volatility shocks received from all other indexes, presented in Figure 14.

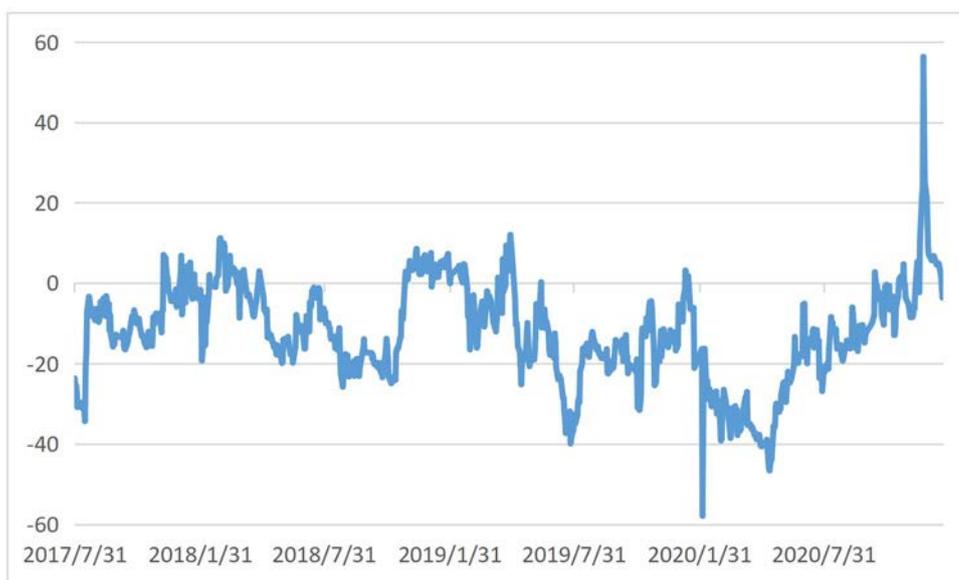

**Figure 14: Net Volatility Spillovers, Ferrous metal futures to six stock indexes**

From Figure 14, the net Volatility Spillovers of ferrous metal futures fluctuate between -60 percent and 60 percent, showing that Ferrous metal futures can be spillover transmitters or recipients. In 2020, under the covid-19 virus situation, it was a spillover recipient at the beginning, then became a spillover transmitter at the end of 2020.

**Net Pairwise Spillovers**

We also calculate rolling-sample net pairwise spillovers between ferrous metal futures and other stock indexes, then present these plots in Figure 15.

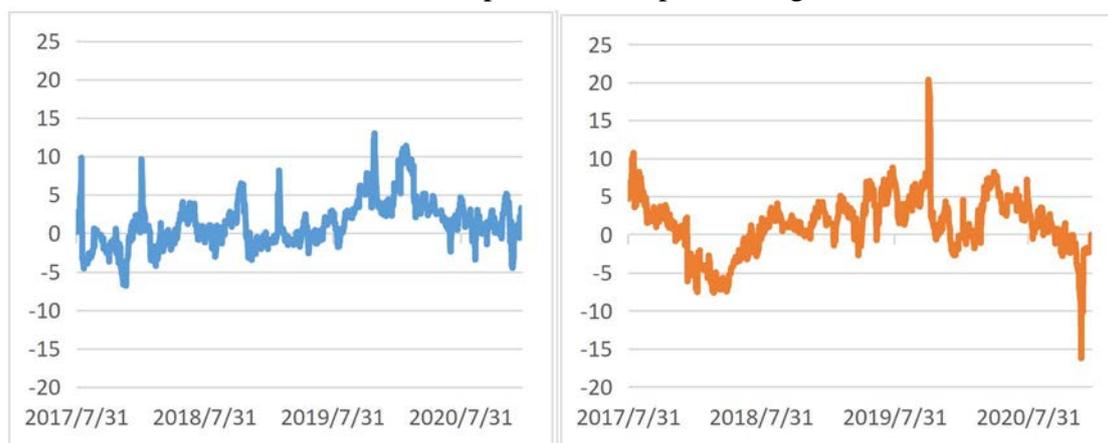

**a) ferrous metal futures-steel industry stock price index    b) ferrous metal futures-home appliance industry index**

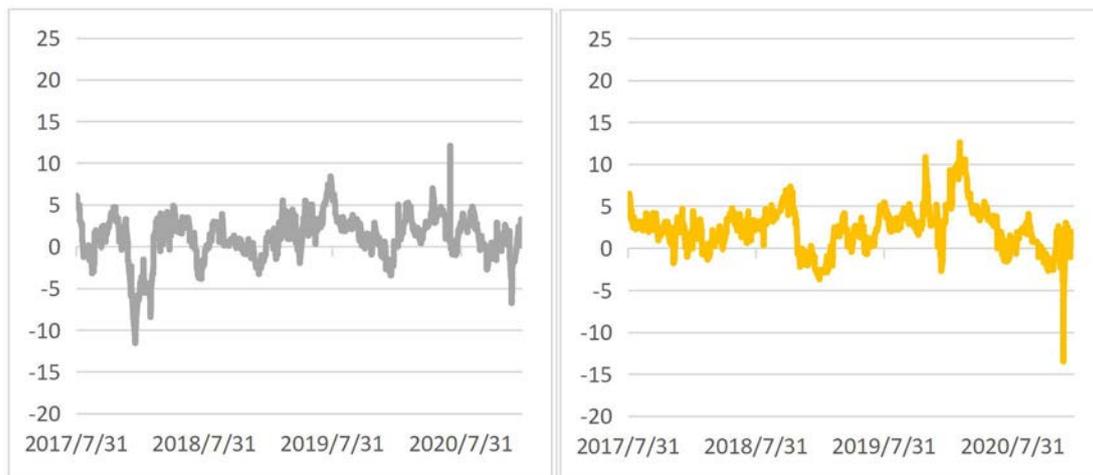

**c) ferrous metal futures-real estate industry stock price index   d) ferrous metal futures-building materials industry stock price index**

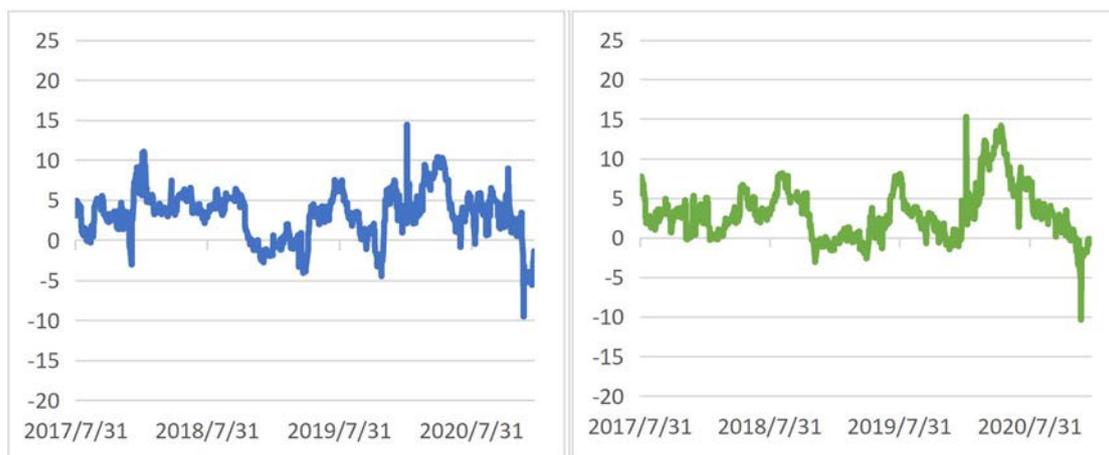

**e) ferrous metal futures-automobile manufacturing industry stock price index**
**f) ferrous metal futures-mechanical equipment industry stock price index**
**Figure 15: Net Pairwise Volatility Spillovers**

From Figure 15, we can find that: 1) ferrous metal futures are time-variant of spillover transmitter or recipient to the steel industry stock price index, the home appliance industry index, the real estate industry stock price index, and the building materials industry stock price index. 2) ferrous metal futures are meanly a spillover transmitter to automobile manufacturing industry stock price index and mechanical equipment industry stock price index. 3) the spillover of ferrous metal futures to each of the other six stock indexes surges at the beginning as a spillover transmitter and then drops down slowly, becoming a spillover recipient during the COVID-19 virus situation in 2020.

**5 Conclusion**

The empirical analysis in this article concludes the conclusions as follows.

Firstly, crude oil, fuel oil, thermal coal, methanol and other energy and chemical products are important industrial raw materials that provide energy for modern manufacturing and industry. The steel industry consumes much energy as an important part of the metallurgical industry. Therefore, the cost of the iron and steel metallurgy industry has an important impact on the price trends of energy and chemical products. Therefore, ferrous metals have a significant transmission effect on energy and chemical future fluctuations. Ferrous metals and nonferrous metals are both important branches of the metallurgical industry. The changes and fluctuations in the futures yields of one of them will be transmitted to the other futures sector, which will produce the same spillover effect. It can be seen that there is a very close volatility spillover relationship between ferrous metal futures and nonferrous metal futures.

Secondly, ferrous metal futures have a significant spillover effect on the stock market's index of steel industry, real estate industry, building materials industry, machinery and equipment industry, and household appliances industry. It shows that the ferrous metal futures market has played an important role in the "barometer" for the Chinese stock market.

Thirdly, ferrous metal futures and other commodity futures, as well as China's stock market, have widespread spillover effects of returns, which shows that the linkage between China's financial markets is becoming stronger and stronger.

Fourthly, the volatility of the ferrous metal futures was low at the beginning of the year 2020 under the COVID-19 virus situation, then raised to a new high level at the end of the year 2020. That is because the demand for metal rises when the economy is recovering from the middle of 2020. Meanwhile, all directional spillover indexes show a reverse trend from the beginning to the end of the year 2020, from a volatility transmitter becoming a recipient or the opposite.

Moreover, the various sectors of the Chinese financial market are not isolated. As an important practice of financial innovation, the futures market has important functions of risk hedging and price discovery. Studying the volatility spillover effect of the ferrous metal futures market can reveal the operating laws of this field and provide ideas and theoretical references for investors to hedge their risks.

We study the volatility spillover effects of the Chinese ferrous metal futures market with the DCC-GARCH model, the BEKK-GARCH model, and the DY(2012) index. However, there are other methods and ideas for describing price fluctuations and spillover effects, such as models based on implied volatility. In the future, it is necessary to carry out related research from other ideas and perspectives.

At the same time, this paper is based on daily trading frequency data. With the rapid development of computer science and technology, high-frequency financial transaction data availability has been greatly improved. In general, the higher the frequency of data, the more market information it contains, which provides a broader way of digging into the information efficiency and spillover effects of the ferrous metal futures market. In future research, high-frequency data based on hours and minutes can be used, and combined with low-frequency data will help us get much more interesting conclusions.

**Date Availability**

The dataset used in this paper is available from the corresponding author upon request.

**Conflicts of Interest**

The authors declare no conflicts of interest.

**References**

Aloui, C., & Nguyen, D.K., & Njeh, H. (2012). Assessing the impacts of oil price fluctuations on stock returns in emerging markets. *Economic Modelling. 29 (6),* 2686-2695.

Seth, N., & Singhania, M. (2018). Volatility in frontier markets: A Multivariate GARCH analysis. J*ournal of Advances in Management Research.* https://doi.org/10.1108/JAMR-02-2018-0017.

Engle, R.F. (1982). Autoregressive Conditional Heteroscedasticity with Estimates of the Variance of United Kingdom Inflation. *Econometrica, 50(4),* 987-1007.

Bollerslev, T. (1986). Generalized autoregressive conditional heteroskedasticity. *Journal of Econometrics, 31(3),* 307-327.

Bollerslev, T., & Engle, R.F., & Wooldridge, J.M. (1988). A capital asset pricing model with time-varying covariances. *The Journal of Political Economy, 96,*116-131.

Campbell, J.Y., & Hamao, Y. (1992). Predictable Stock Returns in the United States and Japan: A Study of Long-Term Capital Market Integration. *Journal of Finance, 1*.

Andersen, T., & Bollerslev, T., & Diebold, F.X., & Labys, P. (1999). Understanding, Optimizing Using and Forecasting Realized Volatility and Correlation. Manuscript, Northwestern University, Duke University and University of Pennsylvania.

Asaturov, K., & Teplova, T., & Hartwell, C. A. (2015). Volatility spillovers and contagion in emerging Europe. *Journal of Applied Economic Sciences,10,*929-945.

Ewing, B.T., & Malik, F., & Ozfidan, O. (2002). Volatility transmission in the oil and natural gas markets. *Energy Economics,24(6),*525-538.

Fung, H.G., & Leung, W.K., & Xu, X.E. (2006). Information Flows Between the U.S. and China Commodity Futures Trading. *Review of Quantitative Finance and Accounting, 21(3),* 267-285.

Lien, D. (2009). Futures Hedging under Disappointment Aversion. *The Journal of Futures Markets, 21(11),* 1029-1042.

Xing, J.P., & Zhou, W.Y., & Ji, F. (2011). Research on information transmission and volatility spillover of stock index futures and spot markets in China. *Securities Market Herald, 2,* 13-19.

Liu, X.B., & Li, Y., & Luo, B. (2012). Research on volatility spillover effects between stock index futures and spot markets. *Macroeconomics, 7,* 80-86.

Kang, S. H., Cheong, C., & Yoon, S. M. (2013). Intraday volatility spillover between spot and futures indices: *Evidence from the Korean stock market. Physica A, 392,*1795–1802.

Sogiakas, Va., & Karathanassis, G. (2015). Informational efficiency and spurious spillover effects between spot and derivatives markets. *Global Finance Journal, 27,* 46-72. http://dx.doi.org/10.1016/j.gfj.2015.04.004

Alotaibi, A.R., & Mishra, A.V. (2015). Global and regional volatility spillovers to GCC stock markets. *Economic Modelling, 45,* 38-49. http://dx.doi.org/10.1016/j.econmod.2014.10.052

Baldi, L., & Peri, M., & Vandone, D. (2016). Stock markets' bubbles burst and volatility spillovers in agricultural commodity markets. *Research in International Business and Finance, 38,* 277–285.

Wang, B.J., & Li, A.W. (2016). Night trading and linkage between China and USA futures markets: from the perspective of volatility spillover effect and dynamic correlation. *Financial Economics Research, 5,* 65-74.

Fu, Q., & Ji, J.W., & Zhong, H.Y. (2017). Continuous trading system and price discovery ability-based on China's gold futures market. *Journal of Applied Statistics and Management, 6,* 1119-1130.

Huo, R., & Ahmed, D.A. (2017). Return and volatility spillovers effects: Evaluating the impact of Shanghai-Hong Kong Stock Connect. *Economic Modelling, 61,* 260-272. http://dx.doi.org/10.1016/j.econmod.2016.09.021

Roy, R.P., & Roy, S.S. (2017). Financial contagion and volatility spillover: An exploration into Indian commodity derivative market. *Economic Modelling, 67,* 368-380. http://dx.doi.org/10.1016/j.econmod.2017.02.019

Dieijen, M., & Borah, A., & Tellis, G., & Franses, P.H. (2018). Big Data Analysis of Volatility Spillovers of Brands across Social Media and Stock Markets. *Industrial Marketing Management, 12,* 1-20. https://doi.org/10.1016/j.indmarman.2018.12.006

Zheng, Y., & Ma, J. (2018). Study on the spillover effect and dynamic relationship between egg futures market and spot market in China. *Journal of China Agricultural University, 23(11),* 222-231. http//doi:10.11841/j.issn.1007-4333.2018.11.23

Reboredo, J. C. . (2018). Green bond and financial markets: co-movement, diversification and price spillover effects. Energy Economics, 74(AUG.), 38-50. http//doi:10.1016/j.eneco.2018.05.030

Chang, C.L., & Liu, C.P., & McAleer, M. (2019). Volatility spillovers for spot, futures, and ETF prices in agriculture and energy. *Energy Economics,81,*779-792. https://doi.org/10.1016/j.eneco.2019.04.017

Li, H.(2020). Volatility spillovers across European stock markets under the uncertainty of Brexit. *Economic Modelling, 84,* 1-12. https://doi.org/10.1016/j.econmod.2019.03.001

Yang, X. (2020). The risk spillovers from the Chinese stock market to major East Asian stock markets: A MSGARCH-EVT-copula approach. *International Review of Economics and Finance, 65,* 173-186. https://doi.org/10.1016/j.iref.2019.10.009

Antonakakis, N., & Floros, C., & Kizys, R. (2016). Dynamic spillover effects in futures markets: UK and US evidence. *International Review of Financial Analysis, 48,* 406-418. http://dx.doi.org/10.1016/j.irfa.2015.03.008

Lau, C.K.M., & Sheng, X. (2018). Inter- and intra-regional analysis on spillover effects across international stock markets. *Research in International Business and Finance, 46,* 420-429. https://doi.org/10.1016/j.ribaf.2018.04.013

Su, X.F. (2019). Measuring extreme risk spillovers across international stock markets: A quantity variance decomposition analysis. *North American Journal of Economics and Finance, 1,* 1-14. https://doi.org/10.1016/j.najef.2019.101098

Kang, S.H., & Uddin, G.S., & Troster, V., & Yoon, S.M. (2019). Directional spillover effects

between ASEAN and world stock markets. *Journal of Multinational Financial Management, 52,* 1-20. https://doi.org/10.1016/j.mulfin.2019.100592

Jiang, Y.H., & Jiang, C., & Nie, H., & Mo, B. (2019) The time-varying linkages between global oil market and China's commodity sectors: Evidence from DCC-GJR-GARCH analyses. *Energy, 166:*577-586. https://doi.org/10.1016/j.energy.2018.10.116

Yin, K., & Liu, Z. & Jin, X. (2020). Interindustry volatility spillover effects in China's stock Market. *Physica A, 539,* 1-15. https://doi.org/10.1016/j.physa.2019.122936

Zhang, W.P., & Zhuang, X.T., & Lu, Y. (2020). Spatial spillover effects and risk contagion around G20 stock markets based on volatility network. *The North American Journal of Economics and Finance,51.* https://doi.org/10.1016/j.najef.2019.101064

Chen, Y.F., & Zheng, B., & Qu, F.(2020)Modeling the nexus of crude oil, new energy and rare earth in China: An asymmetric VAR-BEKK (DCC)-GARCH approach. *Resources Policy,65.* https://doi.org/10.1016/j.resourpol.2019.101545

Diebold, F. X. and K. Yilmaz (2009). Measuring financial asset return and volatility spillovers, with application to global equity markets. The Economic Journal 119 (534), 158-171.

Diebold, F. X. and K. Yilmaz (2012). Better to give than to receive: Predictive directional measurement of volatility spillovers. International Journal of Forecasting 28 (1), 57-66.